# Enhancing Healthcare Recommendation Systems with a Multimodal LLMs-based MOE Architecture

**Jingyu Xu[1*], Yang Wang[2]**

[1]Computing Information Technology, Northern Arizona University, Arizona, U.S, 86011, jyxu01@outlook.com
[2]Cluster BI Inc, Toronto, CA, L5R 2P4, wangyrick100@gmail.com

**Abstract** With the increasing availability of multimodal data, many fields urgently require advanced architectures capable of effectively integrating these diverse data sources to address specific problems. This study proposes a hybrid recommendation model that combines the Mixture of Experts (MOE) framework with large language models to enhance the performance of recommendation systems in the healthcare domain. The MOE framework introduces multiple independent expert models (Experts) to select and activate only a part of the expert models for processing each time the input data is processed, thereby reducing the amount of calculation and improving the professionalism of the processing. We built a small dataset for recommending healthy food based on patient descriptions and evaluated the model's performance on several key metrics, including Precision, Recall, NDCG, and MAP@5. The experimental results show that the hybrid model outperforms the baseline models, which use MOE or large language models individually, in terms of both accuracy and personalized recommendation effectiveness. The paper find mage data provided relatively limited improvement in the performance of the personalized recommendation system, particularly in addressing the cold start problem, Then, the issue of reclassification of images also affected the recommendation results, especially when dealing with low-quality images or changes in the appearance of items, leading to suboptimal performance. The findings provide valuable insights into the development of powerful, scalable, and high-performance recommendation systems, advancing the application of personalized recommendation technologies in real-world domains such as healthcare.

## 1. Introduction
In the digital age, personalized systems referring to information[1] have become important tools for enhancing user experience in the healthcare field, particularly in areas such as health management and wellness advice[2]. With the development of artificial intelligence, some systems have been developed with the aim of providing personalized health and dietary recommendations, and exercise plans based on users' lifestyles, including preferences and historical records[3], thereby improving the efficiency of users' health management and quality of life [4]. While traditional recommendation methods, such as collaborative filtering[5] and content-based filtering[6], are effective in certain contexts, they often face challenges when dealing with sparse data, complex user behaviors, and multiple health factors[7], which limits their application effectiveness in the healthcare sector.

To overcome these limitations, researchers and practitioners have explored hybrid recommendation systems that enhance the accuracy and practicality of recommendations by integrating multiple data sources and advanced learning models. This approach has been widely used in areas with abundant data flow[8], including e-commerce[9], etc. A promising direction is the use of multimodal data with enhanced algorithms[10], which combines users' health records (such as chronic disease history, weight, and dietary records), medical images (such as food images and nutritional composition analysis), and user behavior data. This approach allows for a comprehensive reflection of users' health conditions and lifestyle habits, enabling more precise personalized recommendations. However, handling these heterogeneous data sources, such as how to balance the contributions of different data modalities and how to effectively integrate these data, still needs to be solved.

This study proposes a hybrid model that combines two advanced learning techniques: Mixture of Experts (MOE) [11] and Large Language Model[12]. Hybrid models that combine multiple heterogeneous data and base models have become the mainstream way to improve prediction results[13].The MOE framework is able to share model parameters for related tasks while maintaining specific judging for each task, which gives it an advantage in dealing with complex multi-modal problems[14]. Large Language Model is one of the hottest technologies since the development of AI. It uses deep learning with a large amount of data and performs well in processing unstructured data including images and text. It can provide powerful semantic classification capabilities and efficient learning performance.

By combining these two models, our hybrid approach aims to develop a hybrid recommendation model that can take advantage of the advantages of both, further improving the accuracy of healthy diet recommendations and health advice. The main contributions of this study include:

- A multimodal hybrid recommendation system based on MOE + large language models was developed to enhance the accuracy and personalization of the recommendation system.
- It demonstrates how to integrate text descriptions, image information, and user data to provide a more comprehensive representation of users' health conditions, leading to more precise health and wellness recommendations.
- The model was evaluated using key metrics such as accuracy, recall, NDCG, and MAP@5, and its performance was compared with standalone MOE and LLMs-based models, proving the advantages of the hybrid model.

## 2. Related work

In recommendation systems, the most used traditional methods include collaborative filtering and machine learning-based approaches. Collaborative filtering[15] involves analyzing the similarities in behavior or features between similar users or items to predict user preferences. This method is generally divided into user-based and item-based approaches. However, these methods require large amounts of user data to generate reliable recommendations, and in the healthcare domain, issues such as data sparsity and the cold start problem are particularly severe[16].When recommending for new users or newly emerging health conditions, the lack of sufficient historical data can significantly impact the effectiveness of the recommendation system. Therefore, the medical field itself focuses on how to use advanced technologies to achieve new breakthroughs in heterogeneous data. With the development of artificial intelligence, machine learning-based recommendation systems have gradually been applied, particularly algorithms such as neural matrix factorization, which optimize recommendation performance by modeling the interactions between users and items. Neural matrix factorization, which shown in Figure 1, takes the interaction between user and item as input features, converts it into a one-hot encoding binary sparse vector, user-item pairing table, and finally gives it to the neural network for final output processing.

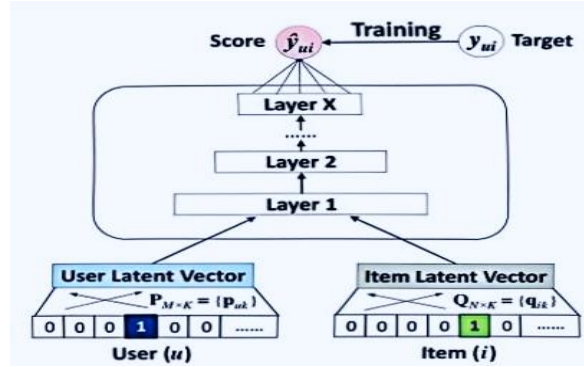

**Figure 1.** Structure of the neural matrix factorization.

However, these methods still struggle to fully capture the dynamic health needs of patients, especially in complex medical environments. Deep learning has made significant progress as a powerful tool in the field of recommendation systems. Deep neural networks are capable of learning nonlinear patterns from high-dimensional data, making them particularly suitable for handling the complex relationships between users and products in the healthcare domain. Pretrained language models, such as BERT [17], have been widely applied in text-based recommendation systems. In the healthcare field, BERT is used together with other algorithms to process data such as medical records, health advice texts, and patient feedback to generate high-quality semantic embeddings for in-depth research on medical human-computer interaction[18]. Meanwhile, the large model of image class represented by Visual Transformer[19] (ViT) has been widely used in medical image analysis, including tasks such as tumor detection and disease prediction. ViT is a new type of computer vision model that shows superior performance in tasks such as image classification and object detection by segmenting the image into multiple small blocks and linearly embedding these blocks into a sequence. As single models gradually reach their capacity limits, researchers have begun to explore recommendation system architectures that can integrate these multi-source data. or example, Luo et al. explored the use of multiple large models to fuse multimodal data [20].Gao et al. summarized some neural network modules that can be used in multimodal heterogeneous data [21]. It is clear that multimodal recommendation systems are a means to more comprehensively describe patients' health status and needs by combining text, image and behavioral data. However, how to integrate and align them and extract effective data insights from them is still a huge problem. The MOE architecture has become a possible heterogeneous data fusion mode. This study aims to fill the gap in practice by combining the MOE architecture and a hybrid multimodal model with a large language model. It provides a valuable framework for the practical application of medical recommendation systems.

## 3. Methodology
### 3.1 Dataset Description
To better align with real-world applications, this study created a small-scale, self-built dataset called "Health Food Recommendations." This dataset contains rich multimodal data, including user demographics, user descriptions, product descriptions, and product images. A total of 177 data entries were collected. The user data primarily includes fields such as query date, a unique user identifier, gender, age, education level, weight, user self-description, and user-uploaded images. The product data includes descriptions and recommendations for health foods, including a unique product identifier, product description, and product category (e.g., food, fruit, or recommendation). All data in the dataset were generated through expert processing of keywords from dietary advice provided by professional doctors based on the patients' conditions.

3.2 Data Preprocessing and Integration

To ensure the quality of input data for the recommendation model, the following preprocessing steps were performed:

- Dataset Merging: The image and text datasets were merged based on common fields, integrating product descriptions and image paths.
- Missing Value Handling: Missing product descriptions were filled with default values, and entries lacking corresponding image mappings were excluded to ensure data consistency.
- Label Mapping: The health food IDs were mapped to numerical labels to facilitate classification and training of the recommendation model.
- Data Splitting: The merged dataset was randomly split into 80% training data and 20% testing data to evaluate the model's generalization ability.

3.3 Model Architecture

Our model is shown in Figure 1. In general, the multimodal data of users and items will first be self-fused, then spliced and aligned with user information, and then sent to the MOE module for training and output of the final recommendation results. The components of our model are described in detail as follows.

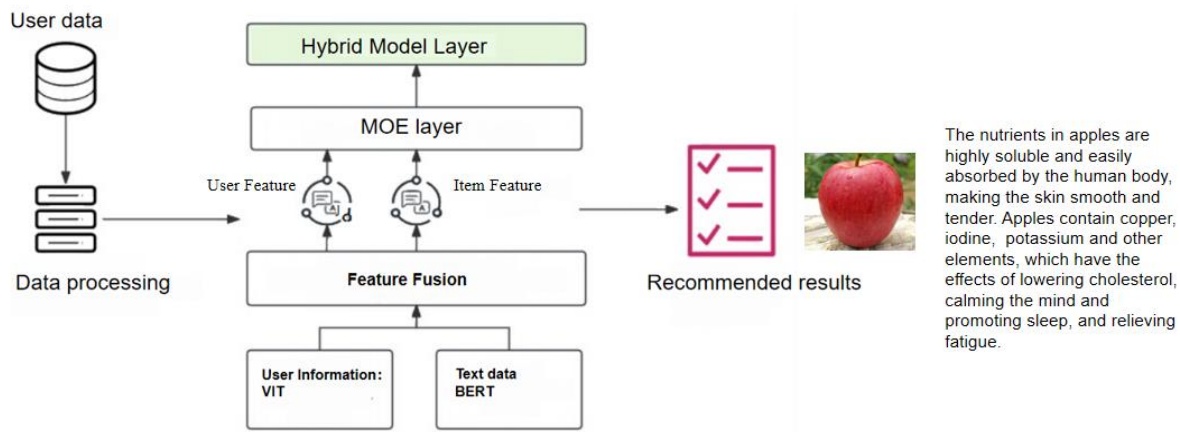

**Figure 2.** Structure of The hybrid recommender system Model.

3.3.1 VIT and BERT

We chose this framework to use BERT and ViT as the basic model, and its model is shown in Figure 3. We used the pre-trained BERT model to process text data. BERT is a classic natural language processing model that combines word vectors, segment vectors, and position vectors to create a comprehensive embedding representation. Word vectors map each word to a high-dimensional space, while segment vectors process individual text pairs. Position vectors provide sequential information to mark the logical sequence of text [22]. We used CLS as the final hidden state as the representation of the entire text sequence. For the image diagnosis part, we used ViT, which converts the visual task into a sequence modeling problem. The principle has been introduced in the introduction part, so we will not go into details. We also use the CLS label in the image to represent the meaning of the image. Afterwards, the representations of the two modalities are concatenated before feature fusion, and then concatenated with the structural data of the user data[23],which is a simple but effective way to get Comprehensive representation.

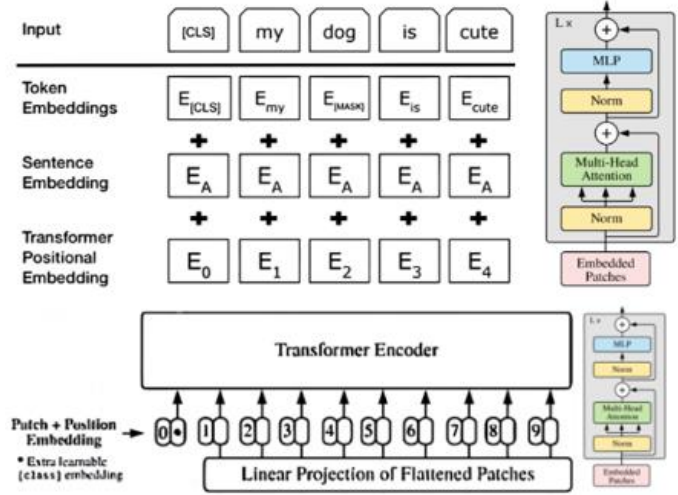
**Figure 3.** Structure of BERT and ViT.

3.3.2 Mixture of Experts (MOE)
  MOE is a multi-task learning framework that allows multiple tasks to share part of the network parameters while maintaining task-specific outputs. This architecture is particularly suitable for scenarios requiring simultaneous learning of multiple related tasks, such as classification and recommendation. Its structure includes the following components:
- Neural Collaborative Filtering Layer: The basic neural network layer is used to perform neural network collaborative filtering on the previously mixed vectorization and extract the common features shared by all tasks from the input data. Finally, this layer compresses the data into 648 dimensions for output.
- Expert Layer: Multiple sub-networks or "experts" specialized in learning specific patterns. Each expert independently processes the shared input and generates its own output. This study used three sets of three-layer transformer blocks as experts. Each transformer[24] layer consists of a self-attention mechanism and a feed-forward neural network stacked together.
- Gating Layer: A task-specific gating mechanism that controls how different experts' outputs are combined, ensuring that each task can dynamically select the most relevant experts based on the input data. In this study, the gating layer was implemented using a stacked model combining a feed-forward neural network and XGBoost[25]. Specifically, XGBoost was used as a base learner for initial training, and its prediction results were fed into the neural network for final predictions.
- Task Head Layer: Each task has a separate head that generates prediction results using the combined expert outputs. In this study, the MOE model generated predictions for multiple tasks, including health food recommendations and health food descriptions.

3.4 Evaluation Matrix

The paper evaluates the performance of models using four indices: Precision@K and NDCG@K. In Formula 1 and 2, Precision@K measures the proportion of relevant items in the top K recommendations to the total number of recommendations to evaluate the recommendation system. It is a variant of Precision. U represents users, $R_u$ refer to the interaction between the user and the item, and $R_u \cap \widehat{R_{u,K}}$ is the result of recommended items the user put the relatively ranking.

$$Precision@K = \frac{1}{|U|} \sum_{u \in U} \frac{|R_u \cap \widehat{R_{u,K}}|}{K} \qquad (1)$$

Normalized discounted cumulative gain (NDCG) is a very common and important metric for evaluating the relevance and ranking of results in recommender systems. DCG@K evaluates the cumulative gain of users for the first K recommended items to calculate the cash-in. IDCG@K represents the maximum possible value of DCG@K under theoretical conditions. Weighting methods are used to ensure that higher-ranked relevance gain greater importance, thereby adjusting the overall ranking. CG calculates the total relevance of recommendations without considering their position, while DCG prioritizes the impact of position loss in the recommendation results for actual situations.

$$NDCG@K = \frac{1}{|U|} \sum_{u \in U} \frac{DCG@K}{IDCG@K} \tag{2}$$

In the recommendation system, MAP@K (Mean Average Precision at K) is used to measure the accuracy of the recommendation system for the top K items in each recommendation list，showing in Formula 3. It is a common method to analyze how many of the recommended items are liked or interested in by the user. For each recommendation, the precision is calculated at the ranking position in the recommendation list and accumulated based on the actual situation. Then, the average precision (equal to AP) is calculated for each user.

$$MAP@K = \frac{1}{N} \sum_{i=1}^{N} AP(i) \tag{3}$$

## 4. Result

The experiment describes the model training process in an optimized computing environment, aiming to combine multi-task learning and the large model techniques of VIR and BERT to improve the performance of personalized recommendation systems. The hardware configuration used in the experiment includes AMD R7 5800H processor, NVIDIA T4 GPU (16GB VRAM), 128GB RAM and 1TB SSD. In terms of software environment, the Colab platform provided by Google was used, combined with common tools such as Python 3.8.10, PyTorch 1.9.0, Transformers 4.10.0. The K in the evaluation method is 5. User data was converted into structured numerical features. The dataset was randomly divided into 80% training set and 20% test set to ensure the fairness of model evaluation. In the feature fusion stage, direct concatenation is used. However, the difference is that the single-modal data output is 648 dimensions, while the multi-modal concatenated data is 1296 dimensions. To ensure consistency, the gating unit will compress the multi-modal data to the original dimension through the feedforward neural network

In the experiment, this study conducted two ablation experiments, aiming to study the performance and function of the model from two aspects to better understand the internal principles. First, this study compared the impact of different modalities and components on the performance of the recommendation system. Secondly, this study also ablated the internal modules of the MOE component in more detail- this is the contribution of this article, and the results are shown in Tables 1 and 2. All training epochs are 100 rounds. The ADAM algorithm[26] is used for optimization.

**Table 1.** Comparison of Models Across metrics

| ID | Model | Precision@K | NDCG | MAP@5 |
|---|---|---|---|---|
| 1 | Text Data | 0.66 | 0.61 | 0.21 |
| 2 | Image data | 0.57 | 0.61 | 0.14 |
| 3 | Multi-modal Data | 0.69 | 0.68 | 0.21 |
| 4 | MOE + Multi-modal Data | 0.73 | 0.71 | 0.24 |

**Table 2.** Comparison of MOE structure

| ID | Model | Precision@K | NDCG | MAP@5 |
|---|---|---|---|---|
| 1 | MOE- Transformers - Stacking[27] | 0.73 | 0.81 | 0.24 |
| 2 | MOE-DNN [28]- Stacking | 0.71 | 0.79 | 0.20 |
| 3 | MOE-CNN[29]- Stacking | 0.70 | 0.77 | 0.19 |
| 4 | MOE- Transformers-DNN | 0.73 | 0.81 | 0.23 |
| 6 | MOE-DNN- DNN | 0.68 | 0.76 | 0.19 |
| 7 | MOE-CNN- DNN | 0.68 | 0.78 | 0.18 |

## 5. Analysis

The experimental results indicate that the multimodal data model performs the best across all key metrics, demonstrating the practical potential of our proposed model in healthcare personalized recommendation systems. By utilizing text and image data extracted from BERT and ViT models, the model shows significant improvements in recommendation accuracy, recall, NDCG, and MAP@5.

The experiment reveals that the model using multimodal data excels in precision (Precision@K). Compared to models that rely solely on text or image data, the inclusion of multimodal data improves the model's precision from 0.66 to 0.69. This suggests that combining text and visual information significantly enhances the relevance of the recommendation results. Specifically, when Precision@K is set to 5, the MOE model with multimodal data achieves a score of 0.73, surpassing the precision of single-modality models, further validating that multimodal data enriches personalized recommendations. In terms of recall, the MOE model with multimodal data also performs better, achieving a recall of 0.81, compared to 0.73 for the text-only model and 0.68 for the image-only model. This indicates that the integration of multimodal data not only increases recommendation accuracy but also covers a broader range of relevant items, ensuring that the system provides users with a wide variety of pertinent and highly relevant content. This result is consistent with the NDCG scores, which also show an improvement in ranking quality due to the fusion of multimodal data. With the text features extracted by BERT and image features from ViT, the relevant items in the recommendation list are effectively ranked higher, leading to higher user satisfaction. The model's NDCG score reaches 0.81, significantly higher than the scores for text-only (0.61) and image-only (0.63) models. The MAP@5 score of 0.24 indicates that the fusion of multimodal data greatly improves the relevance of the top five recommended items, enhancing the practical application value of the personalized recommendation system. In contrast, the single-modality models show lower MAP@5 scores, failing to fully leverage the complementarity of different modalities.

In the ablation experiments on the components of the MOE, we found that, as shown in Table 2, models incorporating the Transformer architecture (such as MOE-Transformers-Stacking and MOE-Transformers-DNN) performed the best, especially in NDCG and MAP@5 metrics. This highlights the powerful contextual learning ability of Transformers in recommendation systems. In comparison, models based on DNN and CNN architectures showed less satisfactory performance, particularly in recommendation ranking and the relevance of the top five items. This confirms that Stacking techniques effectively enhance performance, particularly when multimodal data fusion is involved, significantly improving the overall performance of the recommendation system. This might be because Transformers inherently include feed-forward neural networks, and the marginal effects of deepening the network structure gradually diminish. Some variant architectures, including CNN and Transformer models, theoretically play a certain regulatory role, leading to the overall superiority of the deep neural network-based models.

## 6. Discussion

Experimental evidence confirms that boosting algorithms positively influence the components of the MOE, further enhancing the model's performance. Moreover, the experiment revealed several issues worth further discussion. First, it was found that image data contributed much less to the performance improvement of the personalized recommendation system compared to text data.

We re-applied some mainstream visual model architectures aimed at enhancing image processing, but, as shown in Table 3, the improvements were not significant.

One of the primary issues is the cold start problem. In personalized recommendation systems, the cold start problem occurs when new users or items are introduced, and the model is unable to generate effective personalized recommendations due to the lack of sufficient historical data. Although multimodal data can enhance the quality of recommendations for existing users and items, it is challenging for the model to rely on visual information to make effective predictions for new users or items, especially when image data is unavailable. Therefore, the exploratory search-based approach should be further studied in solving the cold start problem: through exploratory interactions in a visual way [30] or observing user behavior data, user portraits can be better modeled and learned[31].

Another issue concerns image reclassification. The image reclassification problem refers to situations where the model attempts to match images to existing labels or categories, but mismatches may occur, especially when the image quality is poor, or the appearance of the item has changed. In our experiments, despite using various advanced visual models to extract image features, the diversity and complexity of the images still led to unsatisfactory performance in some specific cases. A closer inspection of the user-uploaded images revealed that users often uploaded photos from different application scenarios, including cases, CT scans, web pages, or personal photos. Photos from different environments may require massive amounts of pre-trained data to properly process, which led to unsatisfactory classification results, especially for food items. Therefore, some methods based on special strategies[32] should also be considered for use in combination with heterogeneous data. This also indicates that image data, especially when taken from varied contexts or with lower quality, poses significant challenges for the recommendation system. The model's reliance on visual features is limited, especially when the image content is complex or inconsistent, and it may not fully capture the diversity required for accurate food classification. In addition to improving the model, further implementing certain strategies to fine-tune the performance of LLMs[32] may be another efficient solution.

**Table 3**. Comparison of Multimodal structure

| ID | Model | Precision@K | NDCG | MAP@5 |
|---|---|---|---|---|
| 1 | VIT+BERT | 0.73 | 0.81 | 0.24 |
| 2 | ResNET+BERT | 0.70 | 0.79 | 0.24 |
| 3 | Faster R-CNN+BERT | 0.73 | 0.77 | 0.23 |
| 4 | BEiT+BERT | 0.74 | 0.80 | 0.24 |

## 7. Conclusion

In this work, we propose a hybrid recommendation model based on the MOE framework combined with a large language model, aiming to improve the accuracy and practicality of personalized recommendation systems in the health field. Experimental results show that the hybrid model outperforms the baseline model using only MOE or LLM in multiple key indicators, showing its potential in processing multimodal data, improving recommendation accuracy and personalization effects. Some issues that need to be further addressed were also found in the experiment. The improvement effect of image data in personalized recommendation systems is relatively limited, especially in the cold start problem. In addition, the image reclassification problem also affects the accuracy of the recommendation results when the image is low-quality or the appearance of the item changes, resulting in the system performance failing to reach the optimal level. This study provides valuable insights for the development of efficient and scalable recommendation systems and opens up new directions for the promotion of personalized recommendation technology in real-world applications, especially in the field of health management. Future research can further optimize the processing methods of image data and explore how to effectively integrate different types of data to improve the application effect of the system in various scenarios.